\documentclass[twoside,fleqn]{article}
\usepackage{espcrc2}
\usepackage{graphicx}

\title{
\vspace{-1.0cm}
{\sf \small \rightline{IFIC/02-43, FTUV/02-0930}}
\bigskip
{\Large  NNLO $\tau^+\tau^-$ production cross section at threshold}}
\author{P. Ruiz-Femen\'\i a$^{\mbox{\footnotesize a}}$
\thanks{Talk given at the High-Energy Physics
International Conference in Quantum Chromodynamics 
(QCD 02), Montpellier, France, 2--9 July 2002.} and A. Pich
\address{
       {\em Departament de F\'\i sica Te\'orica, IFIC, 
       Universitat de Val\`encia - CSIC,} \\
       {\em Apartat Correus 22085, E-46071 Val\`encia, Spain}}
       } 
\begin{document}

\newcommand{\nn}{\nonumber}
\newcommand{\mev}{\mbox{\rm MeV}}
\newcommand{\gev}{\mbox{\rm GeV}}
\newcommand{\eqn}[1]{(\ref{#1})}
\newcommand{\MSb}{{\overline{MS}}}
\newcommand{\ep}{\epsilon}
\newcommand{\IM}{\mbox{\rm Im}}
\newcommand{\lsim}{\stackrel{<}{_\sim}}
\newcommand{\gsim}{\stackrel{>}{_\sim}}

\begin{abstract}
The threshold behaviour of the
cross section $\sigma(e^+e^-\to\tau^+\tau^-)$ is analysed,
taking into account the known higher--order corrections.
At present, this observable can be determined to
next-to-next-to-leading order (NNLO) in a combined expansion in powers of
$\alpha_s$ and fermion velocities.
\end{abstract}

\maketitle

\newcommand{\jhep}[3]{{\it JHEP }{\bf #1} (#2) #3}
\newcommand{\nc}[3]{{\it Nuovo Cim. }{\bf #1} (#2) #3}
\newcommand{\npb}[3]{{\it Nucl. Phys. }{\bf B #1} (#2) #3}
\newcommand{\npps}[3]{{\it Nucl. Phys. }{\bf #1} {\it(Proc. Suppl.)} (#2) #3}
\newcommand{\plb}[3]{{\it Phys. Lett. }{\bf B #1} (#2) #3}
\newcommand{\pr}[3]{{\it Phys. Rev. }{\bf #1} (#2) #3}
\newcommand{\prd}[3]{{\it Phys. Rev. }{\bf D #1} (#2) #3}
\newcommand{\prl}[3]{{\it Phys. Rev. Lett. }{\bf #1} (#2) #3}
\newcommand{\prep}[3]{{\it Phys. Rep. }{\bf #1} (#2) #3}
\newcommand{\zpc}[3]{{\it Z. Physik }{\bf C #1} (#2) #3}
\newcommand{\sjnp}[3]{{\it Sov. J. Nucl. Phys. }{\bf #1} (#2) #3}
\newcommand{\jetp}[3]{{\it Sov. Phys. JETP }{\bf #1} (#2) #3}
\newcommand{\jetpl}[3]{{\it JETP Lett. }{\bf #1} (#2) #3}
\newcommand{\ijmpa}[3]{{\it Int. J. Mod. Phys. }{\bf A #1} (#2) #3}
\newcommand{\hepph}[1]{{\tt hep-ph/#1}} 
\newcommand{\hepth}[1]{{\tt hep-th/#1}} 
\newcommand{\heplat}[1]{{\tt hep-lat/#1}}

%%%%%%%%%%%% Nachher loeschen %%%%%%%%%%%%%%%%%%%%%
% \bibliographystyle{physics}
%%%%%%%%%%%%%%%%%%%%%%%%%%%%%%%%%%%%%%%%%%%%%%%%%%%

%%%%%%%%%%%%%%%%%%%%%%%%%%%%%%%%%%%%%%%%%%%%%%%%%%%%%%%%%%%%%%%%%%%%%%%%%
% Beginning of the paper
%%%%%%%%%%%%%%%%%%%%%%%%%%%%%%%%%%%%%%%%%%%%%%%%%%%%%%%%%%%%%%%%%%%%%%%%%

%%%%%%%%%%%%%%%%%%%  Introduction %%%%%%%%%%%%%%%%%%%%%%%%%%%%%%%%%%%%%%%
\section{Introduction}

The Tau--Charm Factory, a high--luminosity ($\sim 10^{33}\;\mbox{cm}^{-2}\;
\mbox{s}^{-1}$) $e^+e^-$ collider with a centre--of--mass energy
near the $\tau^+\tau^-$ production threshold, has been proposed
as a powerful tool to perform high--precision studies
of the $\tau$ lepton, charm hadrons and the charmonium system
\cite{marbella}.
In recent years, this energy region has been only partially explored
by the Chinese BEBC machine ($\sim 10^{31}\;\mbox{cm}^{-2}\;\mbox{s}^{-1}$).
The possibility to operate the Cornell CESR collider
around the $\tau^+\tau^-$ threshold 
has revived again the interest on Tau--Charm Factory physics.

A precise understanding of the $e^+e^-\to\tau^+\tau^-$ production
cross section near threshold is clearly required. The accurate
experimental analysis of this observable could allow to improve the
present measurement \cite{BES} of the $\tau$ lepton mass.
The cross section $\sigma(e^+e^-\to\tau^+\tau^-)$ has already been 
analysed to ${\cal O} (\alpha^3)$ in refs.~\cite{voloshin},
including a resummation of the leading Coulomb corrections.

The recent development of non-relativistic effective
field theories of QED (NRQED) and QCD (NRQCD)  \cite{lepage}
has allowed an extensive investigation of the threshold production of
heavy flavours at $e^+e^-$ colliders. The threshold $b\bar b$ 
\cite{jamin}
and $t\bar t$ \cite{topprodsummary}
production cross sections have been computed to the
next-to-next-to-leading order (NNLO) in a combined expansion in powers of
$\alpha_s$ and the fermion velocities.
Making appropriate changes, those calculations can be easily applied to
the study of $\tau^+\tau^-$ production \cite{ruiz}. One can then achieve
a theoretical precision better than 0.1\%.
%The perturbative ${\cal O} (\alpha^3)$ and ${\cal O} (\alpha^4)$
%contributions are discussed in Section~\ref{sec:pert}.
%Section~\ref{sec:NRQED} contains the relevant non-relativistic
%corrections at low velocities, generating
%${\cal O} (\alpha^n/v^m)$ effects.
%The photon vacuum--polarization is accounted for in
%Section~\ref{sec:vp}.
%In Section~\ref{sec:ew}, bound states corrections are shown to be 
%negligible.
%The numerical results for the $e^+e^-\to\tau^+\tau^-$ cross section and our
%final conclusions are given in Section~\ref{sec:numerics}. 
%%%%%%%%%%%%%%%%%%%%%%%%%%%% Perturbative calculation %%%%%%%%%%%%%%%%%%%%%%%

\section{Perturbative calculation to ${\cal{O}} (\alpha^4)$}
\label{sec:pert}

At lowest order in QED, the $\tau$ leptons are produced by one-photon exchange
in the s-channel, and the total cross section formula reads
\begin{equation}
\sigma_{\mbox{\tiny $B$}} (e^+\,e^- \to \tau^+ \, \tau^-) ={{2 \pi \, \alpha^2} \over {3s}}\,
v\, (3-v^2)\,,
\label{s0}
\end{equation}
where $v=\sqrt{1-4 m_{\tau}^2/s}$ is the velocity of the final $\tau$ 
leptons in the center-of-mass 
frame of the $e^+ \, e^-$ pair which makes $\sigma_B$ vanish when $v\to 0$. 

Electromagnetic corrections of ${\cal O} (\alpha)$ arise from the
interference between the tree level result and 1-loop amplitudes. 
A factor $\alpha/v$ emerges in the 1-loop final
state interaction between the tau leptons, making the cross section at threshold
finite. Furry's theorem guarantees that contributions to 
$\sigma(e^+e^-\to \tau^+\tau^-)$ coming from initial, intermediate and final
state corrections completely factorize at ${\cal O} (\alpha^3)$, including real
photon emission.

Some undesirable features appear at ${\cal O} (\alpha^4)$: The two-loop
$\tau^+\tau^-\gamma$ vertex develops an $\alpha^2/v^2$ term which makes 
the cross section ill-defined when $v\to0$, and multiple photon production
of tau leptons by box-type diagrams and the non-zero interference of initial and final state radition
spoil exact factorization. However, as it has been recently shown \cite{box},
the squared amplitude of the $e^+e^-\to \tau^+\tau^-$ box diagram is
proportional to $\alpha^4v^2$, and so represents a N$^4$LO correction in the
combined expansion in powers of $\alpha$ and $v$, far beyond the scope of this
analysis. In addition, contributions to the total cross section from
diagrams with real photons emitted from the produced taus can be shown to
begin at N$^3$LO, and factorization remains 
at NNLO. The total cross section can thus be 
written as an integration over
the product of separate pieces including initial, intermediate and
final state corrections:
\begin{equation}
\sigma(s)=\int^{s} \, F(s,w)\,\bigg|\frac{1}{1+e^2\Pi{\mbox{\tiny em}}(w)}
\bigg|^{2}\,
\tilde{\sigma}(w)\,dw \,.
\label{secradR}
\end{equation}
The radiation function $F(s,w)$ \cite{kuraev} describes initial state radiation,
including virtual corrections. 
The integration accounts for the effective energy loss
due to photon emission from the $e^+ \, e^-$ pair, and it
includes the largest corrections coming from the
emission of an arbitrary number of initial photons, which can sizeably
suppress the total cross section. $\tilde{\sigma}$ collects only
final-state interactions between the tau leptons, and it is 
usually written in terms of the tau spectral density 
$R_{\mbox{\tiny $\tau$}}$,
\begin{equation}
\tilde{\sigma}(e^+e^-\to\gamma^*\to \tau^+\tau^-)=
R_{\mbox{\tiny $\tau$}}(s)  
\,\sigma_{pt}\,,
\label{R(s)tau}
\end{equation}
with $\sigma_{pt}\, = \,
\frac{4\pi\alpha^2}{3s}$.
The threshold behaviour of the
total cross section will be ruled by the expansion of
$R_{\mbox{\tiny $\tau$}}$ at low velocities.
  
%%%%%%%%%%%%%%%%%%%%%%%%%%%%%%% Coulomb resummation %%%%%%%%%%%%%%%%%%%%%%%%%
 
\section{Non-Relativistic Corrections: NRQED}
\label{sec:NRQED}

A NNLO calculation of the cross section
in the kinematic region where
$v \sim \alpha$ has to account for all terms proportional
to $v\, (\alpha/v)^n\times[1;\alpha;v;\alpha^2;\alpha v;v^2]$ with
$n=1,2,\dots$ 
The leading divergences (i.e.
$\left( \frac{\alpha}{v} \right)^n$, $n > 1$) are resummed in the 
well-known Sommerfeld factor \cite{sommer}
\begin{equation}
|\Psi_{c,\mbox{\tiny E}}(0)|^2 \, = \, \frac {\alpha\pi/v}{1-\mbox{exp}(-
 \alpha\pi/v)}\,,
\label{culfactor}
\end{equation}
multiplying the Born cross section~(\ref{s0}).
A systematic
way to calculate higher-order corrections in this regime requires the use of 
a simplified
theory which keeps the relevant physics at the scale $Mv \sim M\alpha$,
characteristic of the Coulomb interaction.
NRQED \cite{lepage} was designed precisely for this purpose. It is an effective
field theory of QED at low energies, applicable to fermions in non-relativistic
regimes, i.e. with typical momenta $p/M \sim v \ll 1$. Interactions contained in the
NRQED Lagrangian have a definite velocity counting but
propagators and loop integrations can also generate powers of $v$. With appropriate
counting rules at hand, one can prove that all interactions between the non-relativistic
pair $\tau^+\tau^-$ can be described up to NNLO in terms of time-independent potentials
\cite{labelle}, derived from the low-energy Lagrangian. Therefore,
the low-energy expression of the $\tau$ spectral density is related with the
non-relativistic Green's functions \cite{hoangteubner}:
\begin{equation}
R^{\mbox{\tiny NNLO}}_{\mbox{\tiny $\tau$}}(s)  = 
\frac{6\,\pi}{M^2} \,\, \mbox{Im} \Big( C_1\,
G(E)%{\mbox{\boldmath $0$}};
\,-\frac{4E}{3M} \,
G_c(E)\Big)
\label{R(s)NNLOmain}
\end{equation} 
with $C_1$ a short distance coefficient to be determined by matching full and effective
theory results and $E=m_{\tau}v^2$ the non-relativistic energy.
The details of this derivation can be found
in the Appendix B of \cite{ruiz}.
The Green's function $G$ obeys the Schr\"odinger equation corresponding 
to a two-body system
interacting through potentials derived from ${\cal{L}}_{\mbox{\tiny NRQED}}$
at NNLO:
\begin{eqnarray}
\bigg(\!\! -\frac{{\mbox{\boldmath $\nabla$}}^2}{M} 
\!\!\!\!\!&-&\!\!\!\!\!
\frac{{\mbox{\boldmath $\nabla$}}^4}{4M^3} +
V_{c}({\mbox{\boldmath $r$}}) + V_{\mbox{\tiny BF}}({\mbox{\boldmath $r$}})
+ V_{\mbox{\tiny An}}({\mbox{\boldmath $r$}})
-E\bigg)\nonumber\\
&\times& \!\!\!\!G({\mbox{\boldmath $r$}},{\mbox{\boldmath $r$}^\prime},E)
\, = \,
\delta^{(3)}({\mbox{\boldmath $r$}}-{\mbox{\boldmath $r$}^\prime})
\label{Schrodingerfull}
\end{eqnarray}
The term  $-\frac{{\mbox{\boldmath $\nabla$}}^4}{4M^3}$
is the first relativistic correction to the
kinetic energy. $V_c$ stands for the Coulomb potential with 
${\cal O}(\alpha^2)$ corrections. 
At NNLO, the heavy leptons are only produced in triplet 
S-wave states, so we just need to consider the corresponding projection of the 
Breit-Fermi potential $V_{\mbox{\tiny BF}}$.
Finally, $V_{\mbox{\tiny An}}$ is a NNLO piece
derived from a contact
term in ${\cal{L}}_{\mbox{\tiny NRQED}}$, 
which reproduces the QED tree level
s-channel diagram for the process $\tau^+\tau^-\to \tau^+\tau^-$.

A solution of eq.~(\ref{Schrodingerfull}) must rely on numerical or perturbative
techniques. In the QED case, a significant difference between both approaches is
not expected, being $\alpha$ such a small parameter.
Consequently we follow the perturbative approach, using recent
results \cite{hoangteubner}, 
where 
NLO and NNLO corrections to the Green's
function are calculated analytically, via
Rayleigh-Schr\"odinger time-independent perturbation theory around the
known LO Coulomb Green's function $G_c$. We refer the reader to Appendix C
of \cite{ruiz} for complete expressions of the Green's function corrections.
%%%%%%%%%%%%%%%%%%%%%%%%%% Vacuum polarization %%%%%%%%%%%%%%%%%%%

\section{Vacuum Polarization}
\label{sec:vp}

For a complete NNLO description of $\sigma(e^+e^- \to \tau^+\tau^-)$,
two-loop corrections to the photon propagator should be included.
The light lepton
contributions to the vacuum polarization are the standard
1- and 2-loop perturbative expressions.
For the $\tau$ contribution in the threshold vicinity $q^2 \gsim 4M^2$, 
resummation
of singular terms in the limit $v\to 0$ is again mandatory. 
Under the assumption
$\alpha \sim v$, we need to know NLO contributions to
$\Pi_{\tau}(q^2)$, performing the direct
matching for both real and imaginary parts.

In the hadronic sector, we can relate
the hadronic vacuum polarization with the total cross section
$\sigma(e^+e^-\to\gamma^*\to had)$.
%\begin{equation}
%\Pi_{\mbox{\tiny had}}(s)
%= \frac{s^2}{16\pi^3\alpha^2}\,
%\int^{\infty}_{4m_{\pi}^2}dt\,\,
%\frac{\sigma(e^+e^-\to had)}{t(t-s-i\epsilon)}\,.
%\end{equation}
Below 1~GeV, the electromagnetic
production of hadrons is dominated by the $\rho$ resonance
and its decay to two charged pions. The photon mediated $\pi^+\pi^-$
production cross section is
driven by the pion electromagnetic form factor $F(s)$.
An analytic expression for $F(s)$
was obtained in Ref.~\cite{guerrero}, using Resonance
Chiral Theory and the restrictions
imposed by analyticity and unitarity. The obtained $F(s)$
provides an excellent description of experimental data up to energies
of the order
$s_{\rho} \sim $ 1~GeV$^2$.

For the integration region above
$s_{\rho}$, we use $\mbox{Im}
 \Pi_{\mbox{\tiny had}}$ as calculated from pQCD.
Our simple estimate has been proved \cite{ruiz} to deviate
by less than 5$\%$ for the running of $\alpha$ at the scale
$\sqrt{s}=M_Z$. Considering
that $\Pi_{\mbox{\tiny had}}$ modifies
$\sigma(e^+e^-\to \tau^+\tau^-)$ near threshold by roughly $1\%$, 
our result has a global
uncertainty smaller than $0.1\%$ for the total cross section. 

%%%%%%%%%%%%%%%%%%%%% Negligible contributions %%%%%%%%%%%%%%%%

\section{Bound states effect}
\label{sec:ew}

%Electroweak production of heavy quarks
%including threshold effects
%has already been studied in previous papers \cite{penin2,hoang-teubner2},
%and can be easily incorporated in our
%basic formula (\ref{secradR}). However, the characteristic $m_{\tau}^2/M_Z^2$
%supression of this set of corrections and the small value of the
%neutral-current couplings of the leptons make these terms fully negligible in
%our analysis. 
Green's functions develop energy poles below threshold corresponding
to spin triplet (n$^3S_1$) electromagnetic $\tau^+\tau^-$ bound states.
The small width 
of these bound levels, 
dominated by their $e^+e^-$ decay rate
\begin{equation}
\Gamma_{ee}=\frac{m_{\tau}\alpha^5}{6n^3}\approx 
\frac{6.1\cdot 10^{-3}\,\,\mbox{eV}}{n^3}\,,
\nonumber
\end{equation}
make these states very difficult to be resolved experimentally.
For the same reason, the bound states cannot affect the shape of
the cross section at threshold, contrary to the case of heavy quark threshold
production, where bound states play a crucial role.
%Notice, however,
%that the integrated cross section of production of all these bound states 
%could be comparable to some of the 
%NNLO contributions considered in this work.

%%%%%%%%%%%%%%%%%%%%% Numerical analysis %%%%%%%%%%%%%%%%%%%%%%%%%%%%%%%%%

\section{Numerical analysis for $\sigma(e^+e^- \to \tau^+\tau^-)$}
\label{sec:numerics}

%%%%%%%%%%%%%%%%%%%%%%%%%%%%%%%%%%%%%%%%%%%%%%%%%%%%%%%%%%%%%%%%
\begin{figure}[!tb]
\begin{center}
\hspace*{-0.5cm}
\includegraphics[angle=0,height=5.5cm,width=0.5\textwidth]{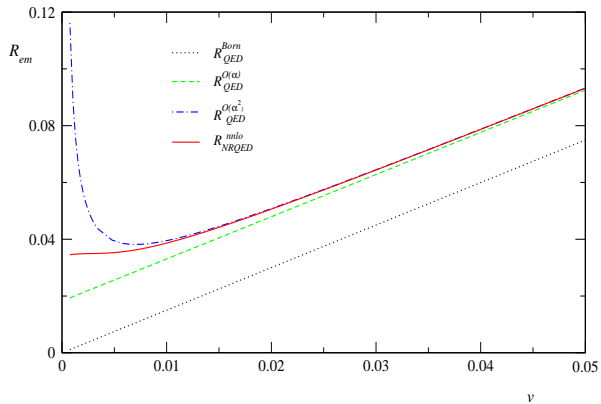}
\end{center}
\vspace*{-1.4cm}
\caption[]{\label{fig:Rthreshold} The spectral density
$R_{\tau}$ at low velocities in both QED and NRQED.}
\vspace*{-.5cm}
\end{figure}
\begin{figure}[!tb]
\includegraphics[angle=0,height=5.5cm,width=0.45\textwidth]{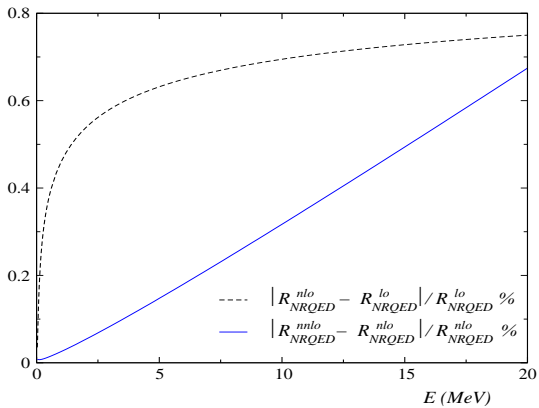}
\vspace*{-1.1cm}
\caption[]{\label{fig:Rsizes} Relative sizes of 
corrections to $R_{\tau}(s)$ as calculated in NRQED.}  
\vspace*{-.7cm}
\end{figure}
\begin{figure}[tb!]
\hspace*{-0.4cm}
\includegraphics[angle=0,height=6cm,width=0.5\textwidth]{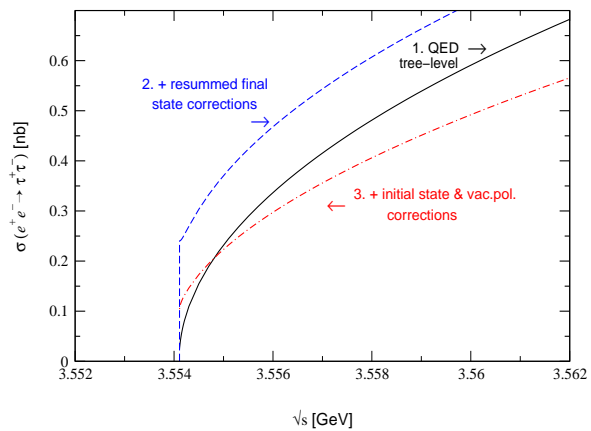}
\vspace*{-1.4cm}
\caption[]{\label{fig:kirk} Total cross section $\sigma(e^+e^-\to
\tau^+\tau^-)$ at threshold.}
\vspace*{-0.7cm}
\end{figure}

%%%%%%%%%%%%%%%%%%%%%%%%%%%%%%%%%%%%%%%%%%%%%%%%%%%%%%%%%%%%%%%%
The need for performing resummations of the leading non-relativistic terms 
$\left( \alpha/v \right)^n [v ,v\alpha,v^2,\dots]$ is evidenced
in Fig.~\ref{fig:Rthreshold}. The QED spectral density vanishes as $v \to 0$,
due to
the phase space velocity in formula (\ref{s0}), which is cancelled by the
first $v^{-1}$ term appearing in the ${\cal O}(\alpha)$ correction, making the
cross section at threshold finite. More singular terms near threshold,
$v^{-2},\dots$
arising in higher-order corrections completely spoil the expected good 
convergence of the QED perturbative series at low $v$. 
This is no longer the case for the effective theory
perturbative series, whose convergence improves as we approach the threshold
point, as shown in Fig.~\ref{fig:Rsizes}, and higher-order corrections
reduce the perturbative uncertainty. In the whole energy range displayed in Fig.~\ref{fig:Rsizes}, the
differences between the NNLO, NLO and LO results are
below 0.8\%, which indicates that the LO result, i.e. the
Sommerfeld factor, contains the relevant physics to describe the threshold
region.
%This is the opposite behaviour to that of the usual
%perturbative QED expansion, 
%Fig.~\ref{fig:Rsizes}b, where the series convergence improves as we move far
%away the threshold.
%%%%%%%%%%%%%%%%%%%%%%%%%%%%%%%%%%%%%%%%%%%%%%%%%%%%%%%%%%%%%%%%%%%%%%%%%
%%%%%%%%%%%%%%%%%%%%%%%%%%%%%%%%%%%%%%%%%%%%%%%%%%%%%%%%%%%%%%%%% 
Adding the intermediate and initial state corrections we have a complete
description of the total cross section of $\tau^+\tau^-$ production, as
shown in Fig.~\ref{fig:kirk}. Coulomb interaction between the produced $\tau$'s,
becomes essential within few
MeV's above the threshold, and its effects have to be taken into account to all
orders. 
Initial state radiation effectively reduces the available center-of-mass energy
for $\tau$ production, lowering in this way the total cross section. 

We should emphasize that NNLO corrections do not modify
the predicted behaviour of the LO and NLO cross section, but are essential to
guarantee that the truncated perturbative series at NLO gets small
corrections from higher-order terms. Hence, we have shown that the
theoretical uncertainty of our analysis of $\sigma(e^+e^- \to \tau^+\tau^-)$ is
lower than 0.1\%. Nevertheless, the statistical
uncertainty of the most recent experiments is still much larger than 
the theoretical one due to low statistics, 
and we should wait for future machines to improve it.

\bigskip \noindent
{\bf Acknowledgements}
I wish to thank S. Narison for the organization of the QCD 02 conference.
This work has been supported in part by TMR, EC Contract
No. ERB FMRX-CT98-0169, by MCYT (Spain) under grant FPA2001-3031, and by ERDF
funds from the European Commission. 

%%%%%%%%%%%%%%%%%%%%%%%%%%%%%%%%%%%%%%%%%%%%%%%%%%%%%%%%%%%%%%%%%%%%%%%%%
%%%%%%%%%%%%%%%%%%%%%%%%%%%%%%%%%%%%%%%%%%%%%%%%%%%%%%%%%%%%%%%%%

% \bibliography{ref}

\end{document}